# Band alignment study of the $Sr_{1-x}Ca_xTaO_2N$ / $H_2O$ interface for photoelectrochemical devices and Hydrogen production.

R. C. Bastidas Briceño,[1,2] V. I. Fernandez,[2,3] and R. E. Alonso,[1,2,4]

[1] *Dto. de Ciencias Básicas, Facultad de Ingeniería Universidad Nacional de La Plata, Argentina.*

[2] *Instituto de Física La Plata, CONICET, Argentina.*

[3] *Dto. de Física, Facultad de Ciencias Exactas, Universidad Nacional de La Plata, Argentina..*

[4] *Instituto de Ingeniería y Agronomía, Universidad Nacional Arturo Jauretche, Argentina.*

**Abstract**

Hydrogen is one of the most promising candidates for clean energy production. But cheap Hydrogen separation and storage is still a big challenge. Photoelectrochemical devices look promising for the decomposition of the water molecule into $2H_2 + O_2$. Every day new materials and combinations are discovered or invented to improve the efficiency of the complex total process. Between them, oxynitrides, like the solid solution $Sr_{(1-x)}Ca_xO_2N$, are good candidates due to the low band gap that lies into the maximum zone of the solar radiation spectrum. A necessary condition for the photoelectrochemical process to work without a bias voltage is that the minimum of the semiconductor conduction band must be more positive than the reduction potential $H^+$ to $H_2$, whereas the maximum of the semiconductor valence band must be more negative than the oxidation potential of $H_2O$ to $O_2$. Thus, band alignment studies in interfaces of semiconductors with water become of great importance. They present several subtleties, as different or simplistic modelling will result in few decimes of eV difference that would lead to a wrong prediction in the alignment. So, to find a trustful method is desirable.

In this work, first-principles calculations based on density–functional theory (DFT) in the all-electron and the pseudo-potential approaches have been performed for the analysis of the band alignment in $Sr_{(1-x)}Ca_xO_2N$ /$H_2O$ interfaces. A detailed study of the modelling of the surface, supercells, and interface with water molecules was done. Experimental data were taken for the structures and photoelectrochemical behaviour and well reproduced by the methodology implemented. Water structures were built from classical molecular dynamics and its electronic structure calculated using DFT. The analysis carried out led to theoretical results compatible with the experimental results: the calculations show that the $Sr_{(1-x)}Ca_xO_2N$ /$H_2O$ is suitable for photoelectrochemical applications, and the partial substitution of Sr by Ca enables the gap and alignment tuning thus enhancing the complex performance.

Keywords: *ab initio* calculations, photoelectrochemical, hydrogen production

## 1- Introduction

The generation of hydrogen emerges as a promising pathway to establish a clean and sustainable energy infrastructure, aiming to decarbonize our current energy economy which heavily relies on fossil fuels [1]. Nowadays, numerous research efforts are directed towards developing more cost-effective and energy-efficient methods for producing this vital resource. For instance, using sunlight to obtain hydrogen from water via photoelectrochemical (PEC) cells based on semiconductors represents a sustainable and environmentally friendly approach [2,3]. Among the necessary conditions and properties for a semiconductor to effectively decompose water molecules into hydrogen and oxygen are: a- photochemical stability: the semiconductor must be stable under illumination and resist photo corrosion in an aqueous solution; b- band edge alignment: the conduction band minimum should be more negative than the hydrogen evolution potential, and the valence band maximum should be more positive than the oxygen evolution potential; c- efficient charge separation: the material should facilitate the separation and migration of photo-generated electron-hole pairs to prevent recombination; d- catalytic activity: the surface of the semiconductor should possess active sites that can efficiently catalyse the water-splitting reactions. These factors, combined with an appropriate band gap, are crucial for the semiconductor's performance in water-splitting applications.

Perovskite oxynitrides of the $ABO_2N$ type exhibit distinctive properties that position them as ideal candidates for prominent roles in photo electrochemistry. Their relevance lies in their energy gap ranges, which oscillate between 1.8 and 2.5 eV, depending on the A and B cations in the crystal structures. This represents efficient absorption of visible light at wavelengths between 500–700 nm [4]. Previous studies have investigated $SrTaO_2N$ as a possible photocatalyst by analysing the band alignment of the semiconductor and water molecules in a $SrTaO_2N/H_2O$ interface. One of these also analysed how variations in the material's lattice constants influence the band-edge alignment concerning the water redox potential [5]. This variation and tuning, in principle, can be effectively done by producing thin films of the materials grown on different substrates with different lattice mismatch.

However, according to experimental studies the photocatalytic efficiency of perovskite oxynitrides has been limited for water splitting, even with the intervention of sacrificial agents. This phenomenon is attributed to critical factors closely related to charge separation, migration, and transfer, such as the presence of recombination centres [6, 7], structural distortions like the tilting of $Ta(O, N)_6$ octahedra [8], and inadequate band edge alignments [6]. Recent advances have been made by strategically introducing Ca into the A-site of the oxynitride [9] or by synthesising a solid solution between $SrTaO_2N$ (STN) and $CaTaO_2N$ (CTN) [10]. These modifications have demonstrated improved efficiency in the water-splitting reaction in these materials.

Computational quantum simulations have proven to be a suitable tool for understanding the microscopic processes that drive the physical properties of materials. Beyond the simplifications that are made to model the systems due to the complexity of real materials (especially in disordered and ceramic materials), they can lead to a better understanding of the individual processes involved.

In this study, the band edge alignment in the solid solution $Sr_{1-x}Ca_xTaO_2N$ (CSTN) will be investigated for the first time for x= 0, 0.25, 0.5, 0.75, and 1, using *ab initio* quantum mechanical models. For each composition, the supercell approach has been implemented to reproduce the different proportions of Ca and Sr. Different details of the modelling were analysed such as: supercell height, vacuum size, atomic relaxation of the atoms at the semiconductor surface, band bending, liquid water simulation, and gap determination. All these factors were crucial to understand the behaviour of the semiconductor surface and its capability to produce water splitting [11].

## 2- Computational Method

*Ab initio* calculations were performed using two different approaches to solve the Kohn-Sham equation within the framework of Density Functional Theory (DFT): (i) the Full-Potential Augmented Plane Wave plus local orbital (FP-APW) method in the scalar relativistic version [12, 13], implemented in the WIEN2k code [14, 15], and (ii) the plane wave and pseudopotential method, implemented in the Quantum Espresso (QE) code [16].

For the calculation of band gaps, WIEN2k was used due to its implementation of the modified Becke-Johnson exchange potential (TB-mBJ) [17] in the TB-mBJ0 [17] and TB-mBJ1 [18] versions. This approach significantly improves the predicted values compared to traditional LDA and GGA calculations. The latter are known for their tendency to underestimate band gap values, as they are accurate only for the ground state. On the other hand, previous studies have shown that the TB-mBJ method provides highly improved results comparable to those obtained by hybrid methods, which are much more computationally expensive.

After convergence studies for the calculations performed with WIEN2k, the following parameters were selected: muffin-tin radii of 2.5 Å for Sr and Ca, 2 Å for Ta, and 1.7 Å for N and O. The parameter $RK_{Max} = R_{MT}*K_{Max}$ was set to 7, where $R_{MT}$ is the smallest muffin-tin radius and $K_{Max}$ is the largest wave number of the basis set. Integration in reciprocal space was performed using the tetrahedron method with 300 k-points in the first Brillouin zone (reduced to 10 k-points in the IBZ).

For the analysis of structural properties and band alignment calculations, the QE pseudopotential method was used with the GGA PBE-Sol exchange-correlation potential. After convergence tests, computation time assessments, and comparison with experimental data, the following selections were made: the ultrasoft pseudopotentials Rappe Rabe Kaxiras Joannopoulos [19] were used for electron-ion interactions; the wave functions were expanded using plane waves with a kinetic energy cutoff of 80 Ry, and a charge density cutoff of 800 Ry. The IBZ was sampled using the Monkhorst-Pack scheme with a 6 × 6 × 1 mesh [20].

3- Modelling the interfaces

i- Construction and optimization of STN, CTN, and their solid solutions structures.

Bulk CTN exhibits an orthorhombic perovskite configuration with space group *Pnma* at room temperature (RT), while bulk STN is tetragonal and belongs to the space group *I4/mcm* at RT. The solid solutions $Sr_{1-x}Ca_xTaO_2N$, at high concentrations, retain the structures of the predominant cation, either Ca or Sr, but exhibit a morphotropic phase transition for 40 < $x$ < 60, as reported by Wang et al. [10]. In that study, the authors found the structures for the compositions $x$ = 0, 0.20, 0.40, 0.60, 0.80, and 1. For the present study, six structures were constructed with the compositions: $x$ = 0, 0.25, 0.5, 0.75, and 1, starting from the unit cells of the pure CTN and STN compounds and making the appropriate cationic substitutions of Ca and Sr for each composition. Subsequently, an interpolation of the experimental data from Wang et al. was performed for each of the constructed compositions, and these values were applied to the thus-formed structures. For the composition x = 0.5, since it was not previously reported and is very close to the morphotropic phase transition, structures were constructed in both the *Pnma* (Fig. 1(a)) and *I4/mcm* (Fig. 1(b)) phases. Then, optimization was carried out using DFT calculations for each composition, relaxing the cell constants and atomic positions. The obtained data are presented in Table 1. In this table we also show the interpolated cell constants and volume of the experimental data at intermediate concentrations. Total energy calculations for $x$ = 0.50 predict that the I4/mcm structure presents a lower energy than the Pnma structure, with an energy difference of 0.118 eV. This difference is small but within the precision limits of DFT calculations (0.001 eV). Consequently, the structures with $x$ = 1 and 0.75 belong to the Pnma space group, while the structures with x = 0.5, 0.25, and 0 should belong to the I4/mcm space group. Although, considering that small content of vacancies and defects in real samples can alter the phase stability, band alignment for the $x$ = 0.5 case will be performed with both space groups for comparison purposes

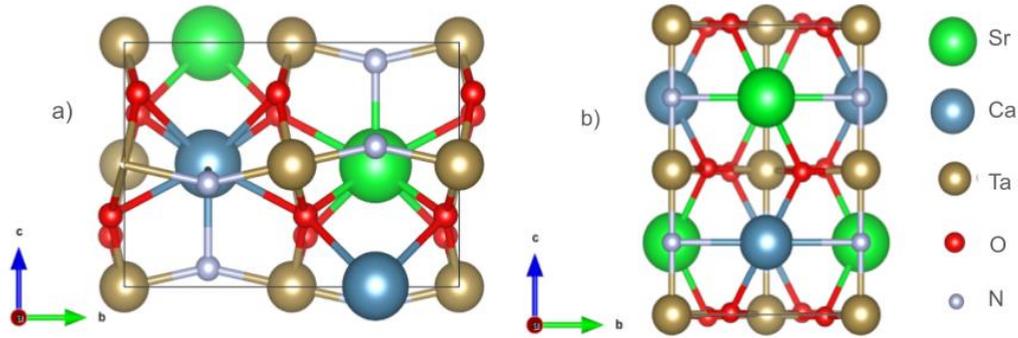

**Figure 1:** Bulk structures of $Sr_{0.5}Ca_{0.5}TaO_2N$: a) *Pnma* b) *I4/mcm*.

**Table 1:** Experimental interpolated and calculated cell constants, volume and T-(O,N) distances for $Sr_{1-x}Ca_xTaO_2N$ solid solutions.

| Composition | Phase | Method | $a$ (Å) | $b$ (Å) | $c$ (Å) | Volume (Å$^3$) | Ta - O (Å) | Ta - O (Å) | Ta - N (Å) | Ref. |
|---|---|---|---|---|---|---|---|---|---|---|
| $x = 1$ | | | 5,6239(3) | 7,8954(4) | 5,5473(3) | 246,32(3) | 2,017 | 2,044 | 2,025 | [10] |
| $x = 0,75$ | *Pnma* | | 5,6070(3) | 7,9385(5) | 5,6366(3) | 250,892(3) | | 2.003(1) | | |
| $x = 0,50$ | | experimental interpolated data | 5,6451(2) | 7,9894(4) | 5,6585(3) | 255,205(2) | | 2.012(2) | | |
| $x = 0,25$ | | | 5,6756(2) | 8,0422(2) | 5,6756(2) | 259,060(2) | | 2,019(2) | | |
| $x = 0$ | *I4mcm* | | 5,7049(3) | 5,7049(3) | 8,0499(5) | 261,99(4) | 2,025 | 2,025 | 2,012 | [10] |
| $x = 1$ | | | 5,502 | 8,123 | 5,511 | 246,260 | 2,020 | 2,021 | 2,0413 | |
| $x = 0,75$ | *Pnma* | | 5,562 | 8,112 | 5,562 | 250,880 | 1,977 | 2,047 | 2,0406 | |
| $x = 0,50$ | | calculated | 5,590 | 8,153 | 5,598 | 255,090 | 1,943 | 2,088 | 2,042 | |
| $x = 0,50$ | | | 5,588 | 8,155 | 5,599 | 255,147 | 2,023 | 2,023 | 2,842 | |
| $x = 0,25$ | *I4mcm* | | 5,603 | 5,603 | 8,216 | 257,890 | 2,024 | 2,028 | 2,050 | |
| $x = 0$ | | | 5,644 | 5,644 | 8,222 | 261,900 | 2,029 | 2,029 | 2,055 | |

**ii- Interfaces and optimization**

To construct a semiconductor/water interface, the first step is to create a semiconductor surface. Since most programs used to calculate electronic structure employ periodic boundary conditions, this is achieved by intercalating a vacuum layer over the unit cell structure in a selected direction. This effectively forms a periodic structure of semiconductor/vacuum/semiconductor... in that direction. The vacuum layer is then shaped like a box, into which water molecules will be included later (Fig. 2), as described in the following subsection. The semiconductor layer, like the vacuum layer, is also confined in the same direction. Therefore, it is crucial to give the semiconductor layer sufficient thickness to ensure that surface effects do not dominate over bulk effects, thereby accurately simulating a real compound. To achieve this, the properties of the bulk solid must be reproduced in the central region of the semiconductor layer, provided it is thick enough. This increase in thickness linearly increases the number of atoms in the system, which in turn increases the computation time by approximately the number of atoms raised to the third power. Thus, in this work, before constructing the semiconductor/$H_2O$ interfaces, a convergence test of the number of the semiconductor unit cells in the surface direction was performed to establish the correct and minimum size of the supercells (SCs). Since the relevant magnitude for the calculations performed (see below) is the Hartree potential (HP), it was used to analyse the difference between the average values of the HP potential on the semiconductor side, ensuring that this difference does not vary significantly with increasing thickness. The QE code was used for this purpose. For the orthorhombic $CaTaO_2N$ structure, an optimal number $n = 5$ of layers was obtained. For the tetragonal $SrTaO_2N$ structure, the optimal $n = 3$ was found. Considering these $n$, two types of SCs were constructed: 2x1x5 for orthorhombic structures and 2x2x3 for the corresponding tetragonal structures. The dimensions of the structures to be used are presented in Table 2.

**Table 2:** $Sr_{1-x}Ca_xTaO_2N$ supercell constants in Å and number of water molecules into the water box.

| Composition | Phase | 2a | 1b | 5c | m |
|---|---|---|---|---|---|
| $x = 1$ | | 11,004 | 8,123 | 27,555 | 39 |
| $x = 0,75$ | Pnma | 11,124 | 8,112 | 27,810 | 40 |
| $x = 0,50$ | | 11,180 | 8,153 | 27,990 | 40 |
| | | 2a | 2b | 3c | |
| $x = 0,50$ $x = 0,25$ | I4mcm | 11,118 | 11,118 | 24,618 | 54 |
| | | 11,206 | 11,206 | 24,648 | 55 |
| $x = 0$ | | 11,296 | 11,296 | 24,852 | 56 |

### iii-    Modelling Water Boxes

Six water boxes were constructed to be attached to the optimised SCs of the different compositions, as is shown in Fig. 2. The number of water molecules $m$ needed to maintain approximately the density of water under normal pressure and temperature conditions (1 g/cm$^3$) was calculated (Table 2). The water molecules were randomly distributed within the different boxes, with cell constants 2*a* x 1*b* x 13.17 Å for *Pnma* and 2*a* x 2*b* x 13.17 Å for *I4mcm*.

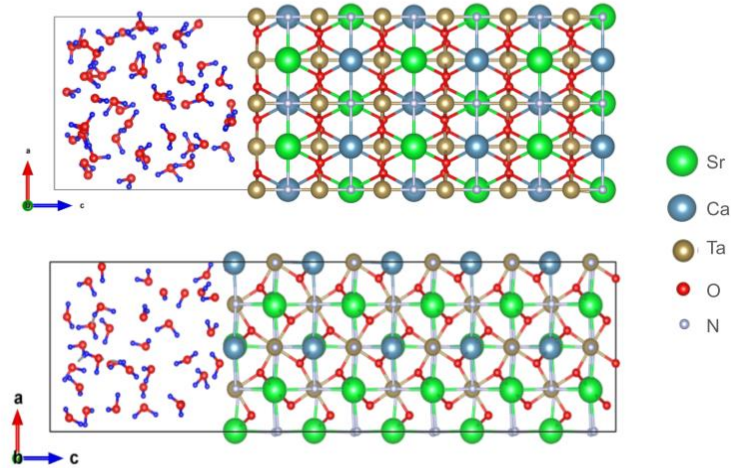

**Figure 2:** CSTN Interface. Tetragonal 2x2x3 (top) and orthorhombic 2x1x5 (bottom).

Classical molecular dynamics (MD) simulations were then performed using the DLPOLY program, based on the work of Y. Wu [11]. The initial configuration consisted of *m* H$_2$O molecules in the box. The interaction between water molecules was modelled using the TIP4P potential within the NVT ensemble for 100 ps at 300 K. The temperature was controlled by a Berendsen thermostat, and the final instantaneous atomic configurations were used as input for the DFT calculations of each box.

In Figure 3, the obtained density of states (DOS) graph for a system of 56 water molecules after this procedure is shown, in good agreement with Fig. 3 in the work of Wu et al. [11].

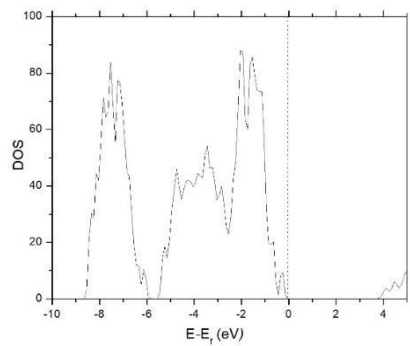

**Figure 3:** Total DOS (in arbitrary units) for a liquid-water system of 56 molecules after 100 ps classical MD plus a final relaxation using DFT.

### 4- Band alignment method:

There are several methods based on DFT for band alignment, and the choice depends on the materials and characteristics of the interface. In [11], Wu, Chan, and Ceder proposed a method to calculate band alignment for semiconductor-water interfaces that considers the band bending at the edge, which is crucial for the alignment of solid solution structures, as shown in Figure 4:

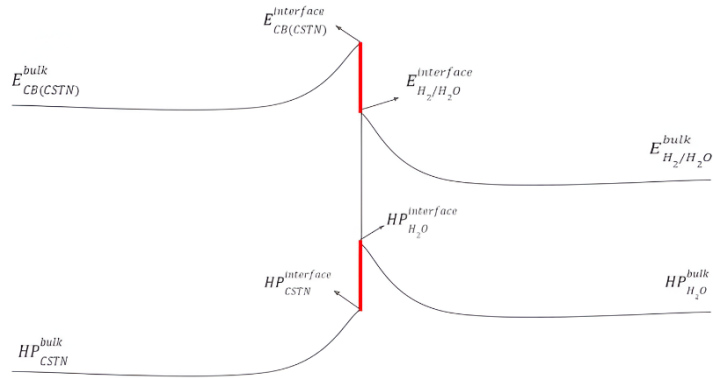

**Figure 4:** Band Alignment Diagram of the SCTN/$H_2O$ interface: On the left side is the CB minimum of the solid solution and the mean Hartree potential, and on the right side are the corresponding levels for $H_2O$.

This procedure involves calculating three terms independently, known as the three-step method for band edge alignment. The first assumption is that all relevant energy levels for the band alignment calculation at the interface are related to the corresponding levels of the SCTN structure and water through band bending at the interface, following the nomenclature of Fig. 4, as described by the following equations:

$$E_{CB(SCTN)}^{interface} - HP_{SCTN}^{interface} = E_{CB(SCTN)}^{bulk} - HP_{SCTN}^{bulk} \quad (1)$$

$$E_{H2/H2O}^{interface} - HP_{H2O}^{interface} = E_{H2/H2O}^{bulk} - HP_{H2O}^{bulk} \quad (2)$$

Based on this assumption, Wu et al. demonstrate that the relative position of the conduction band of a semiconductor and the $H_2O/H_2$ level of water at the interface is given by:

$$E_{CB(SCTN)}^{interface} - E_{H2/H2O}^{interface} = \left(E_{CB(SCTN)}^{bulk} - HP_{SCTN}^{bulk}\right) -$$

$$\left(E_{H2/H2O}^{bulk} - HP_{H2O}^{bulk}\right) +$$

$$\left(HP_{SCTN}^{interface} - HP_{H2O}^{interface}\right) \quad (3)$$

Where: a) the first term on the right side of Eq. 3 accounts for the difference between the energy of the CB of the semiconductor obtained from its initial bulk optimization and the corresponding mean HP; b) the second term represents the $H_2O/H_2$ acceptor level relative to the HP in the water system. To obtain this, it is necessary to calculate the lowest unoccupied molecular level (LUMO) of water, known as the water acceptor level. These calculations can be performed using classical molecular dynamics (MD), replacing a water molecule with a hydronium ion ($H_3O^+$) in the water system; and: c) the third term represents the difference between the average HP in the semiconductor and the water sectors of the interface. For its calculation, the relaxation of the two layers of atoms near the surface and all the atoms in the water boxes for the fourteen SCs constructed was performed. Subsequently, the mean HP for the interface on both the water side and the SCTN was calculated.

A similar procedure is required to determine the alignment of the valence band (VB) with respect to the $O_2$ redox level: $E_{VB(SCTN)}^{interface} - E_{O2/H2O}^{interface}$. In principle, a similar equation to Eq. 3 can be implemented. However, this step is not necessary, as the bands are expected to bend approximately constantly, then:

$$E_{VB(SCTN)}^{interface} = E_{CB(SCTN)}^{interface} - GAP_{SC} \quad (4)$$

$$E_{O2/H2O}^{interface} = E_{H2/H2O}^{interface} - \Delta_{LUMO-HOMO} \quad (5)$$

From these equations, the following equation is applied:

$$E_{VB(SCTN)}^{interface} - E_{O2/H2O}^{interface} = \left(E_{CB(SCTN)}^{interface} - E_{H2/H2O}^{interface}\right) + \Delta_{LUMO-HOMO} - GAP_{SC} \quad (6)$$

The first term on the right side of Eq. 6 is simply the result of Eq. 3. The second term is the redox potential of water (1.23 eV), while the last term is the band gap in the semiconductor. The gap was calculated for each composition using the WIEN2k, as explained below.

5- **Results**

In this work, the alignment of the VB and CB vs. the $H_2O$ redox levels, as well as the gap of the solid solutions for three compositions of the *Pnma* phase and three of *I4/mcm* was studied, using the SC model described above. The calculation of each term of Eq. 3 was performed for each one of the compositions. The results of the calculation of the elements of the first term for each structure are shown in Table 3.

For the second term, we used the calculated value of 0.7 eV from the work of Wu et al. [11]. It is important to note that this value is independent of the system under study and is exclusively determined by the properties of water.

**Table 3:** $E_{CB(SCTN)}^{bulk}$ and $HP_{H2O}^{bulk}$ for the different compositions in the bulk CSTN solutions in eV.

| Phase | Pnma | | | I4mcm | | |
|---|---|---|---|---|---|---|
| Composition | $x = 1$ | $x = 0,75$ | $x = 0,50$ | $x = 0,50$ | $x = 0,25$ | $x = 0$ |
| $E_{CB(SCTN)}^{bulk}$ | 12,25 | 12,38 | 12,45 | 12,80 | 12,74 | 12,58 |
| $HP_{SCTN}^{bulk}$ | 5,43 | 5,40 | 5,39 | 5,43 | 5,41 | 5,36 |

Finally, for the calculation of the third term, it is important to note that the structures on the semiconductor side are polar. The orthorhombic structures result in a surface charge of ±8e, while the tetragonal ones have a (smaller) value of ±2e. Consequently, a uniform electrostatic field is generated in opposite directions on either side of the interface, which adds to the potential of the atomic arrangement. This effect can be observed in Figure 5, which shows the HP obtained on both sides

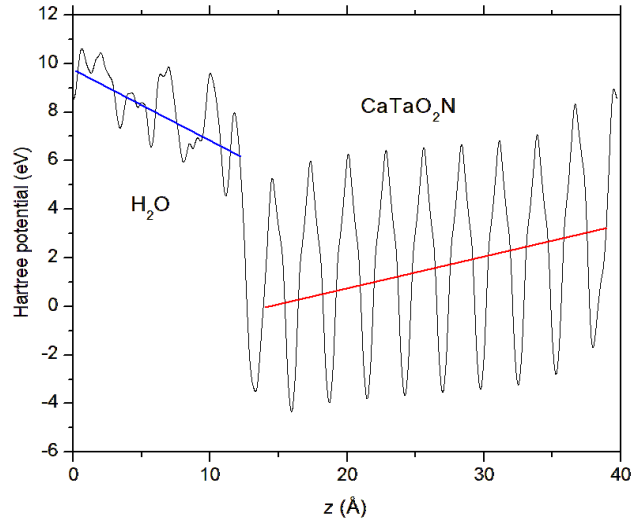

of the CSTN/H$_2$O interface for the x = 1 structure.

**Figure 5:** HP at the CaTaO$_2$N/H$_2$O interface. The blue line on the left represents the average HP of water, and the red line on the right corresponds to the average HP of the SC.

Since water molecules are polar, the presence of the electric field causes a partial rearrangement of them, aligning to some extent the individual dipole moments in the opposite direction of the field, thus producing a screening effect. This entire process has been studied as follows: first, the Hartree potential (HP) was calculated in the already relaxed system where the water molecules were removed, i.e., the water box was replaced with an empty box. Figure 6 shows the comparison of the HPs of both structures. On the water side, the HP corresponding to the semiconductor + water structure is approximately 5.7 eV lower than the HP in the curve corresponding to the semiconductor + vacuum structure, thus illustrating the screening effect.

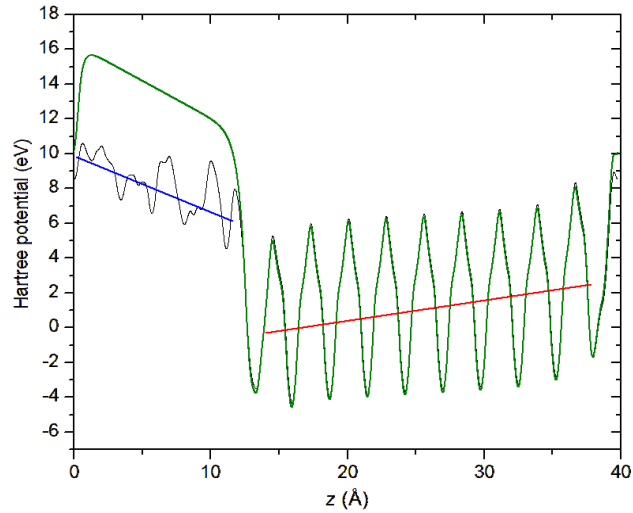

**Figure 6:** HP at the $CaTaO_2N/H_2O$ interface. The blue line on the left represents the average HP line of the water, the red line on the right corresponds to the average HP line on the semiconductor side, and the green line shows the HP obtained in the semiconductor + vacuum cell.

Additionally, the HP calculation was also performed for the structure where the semiconductor atoms were replaced with vacuum, resulting in a water + vacuum structure. Since the initial structure was taken from the already relaxed system, the water molecules remained frozen in their partially polarised position. The effect of this reverse polarisation can be clearly seen in Figure 7. Compared to Fig. 6, the inversion of the HP slopes can be seen on both the water side and the vacuum side. Figures 5, 6 and 7 correspond to orthorhombic structures. The same effect occurs also for the tetragonal structures, but since the net surface charges are four times smaller, the effect is practically imperceptible. In the complete semiconductor + water system, the total field results from the sum of both, with the semiconductor field predominating, as observed in Fig. 8 of the case $x=1$, tetragonal *I4mc*m $SrTaO_2N$. It should be noted that this effect of the HP slopes on either side of the interface does not alter their average values, needed for Eq. 3.

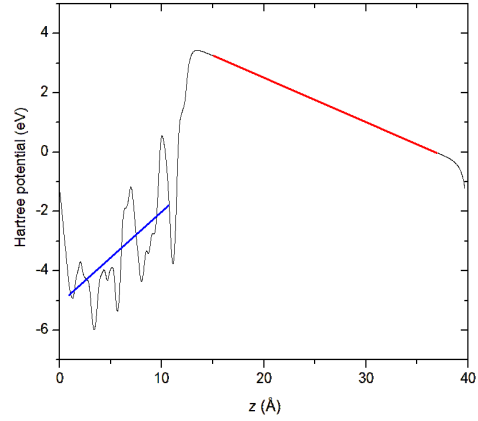

**Figure 7:** HP in the water + vacuum system. The blue line on the left represents the average HP line of the water, and the red line on the right corresponds to the average HP line of the vacuum.

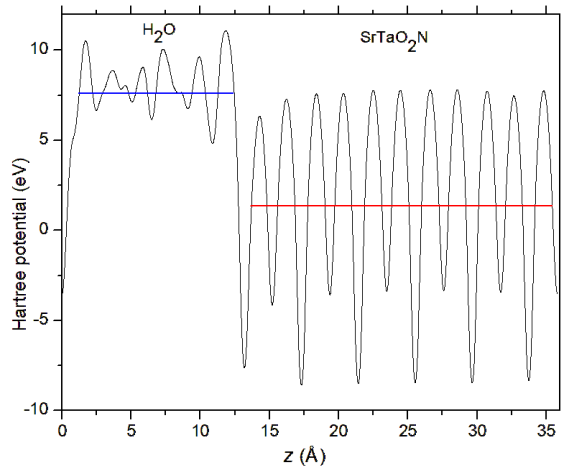

**Figure 8:** Average HP value in the ab plane for the SrTaO₂N/H₂O interface. The blue line represents the average value in the c direction on the water side, and the red line corresponds to the SC side.

Then, the results obtained for the third term of Eq. 3 for all the studied compositions are presented in Table 4. The combination of the data from Table 3 and 4 together with Eq. 3 enable to compute $E_{CB(SCTN)}^{interface} - E_{H2/H2O}^{interface}$ for each structure. The results are presented in Table 5.

**Table 4:** Average HP on each side of the CSTN interface in eV.

| Space group | Pnma | | | I4mcm | | |
|---|---|---|---|---|---|---|
| Composition | x = 1 | x = 0,75 | x = 0,50 | x = 0,50 | x = 0,25 | x = 0 |
| $HP_{SCTN}^{interface}$ | 1,99 | 2,04 | 2,03 | 1.86 | 1,99 | 1,58 |
| $HP_{H2O}^{interface}$ | 8,11 | 7,80 | 8,02 | 7.80 | 7,51 | 7,67 |

**Table 5:** Obtained CB alignment: $E_{CB(SCTN)}^{interface} - E_{H2/H2O}^{interface}$ in eV for each composition for the SCTN/H$_2$O interfaces.

| Space group | Pnma | | | I4mcm | | |
|---|---|---|---|---|---|---|
| Composition | x = 1 | x = 0,75 | x = 0,50 | x = 0,50 | x = 0,25 | x = 0 |
| $E_{CB(SCTN)}^{interface} - E_{H2/H2O}^{interface}$ | 0.0 | 0.53 | 0.36 | 0.73 | 1.11 | 0.70 |

Finally, to also obtain the maximum of the valence band for alignment analysis, the gap of each composition was calculated to use them in Eq. 6. To compare experimental and calculated results, an interpolation adjustment of the former was performed, as shown in Fig. 9, using data from reference [10]. This is due to the measurements in that study were performed for compositions different from those in this study. The calculated values were obtained using PBE, as well as different parameterizations of TB-mBJ, and are presented in Table 6.

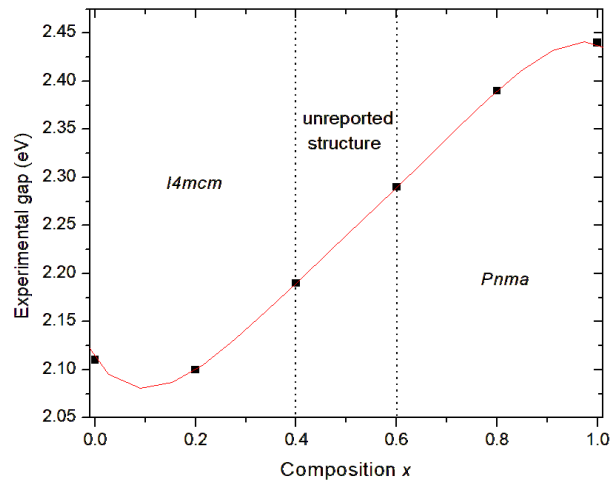

**Figure 9:** Experimental gap vs. composition fitted with a 5th-degree polynomial

**Table 6:** CSTN gap values for different compositions: extrapolated/experimental and theoretical values calculated with PBE, TB-mBJ0, and TB-mBJ1.

| Composition | Space group | PBE (eV) | TB-mBJ0 (eV) | TB-mBJ1(eV) | Gap (eV) |
|---|---|---|---|---|---|
| x = 1 |  | 0,715 | 2,105 | 2,244 | 2,44 [10] |
| x = 0,75 | Pnma | 1,137 | 2,229 | 2,335 | 2,37 |
| x = 0,50 |  | 1,128 | 2,186 | 2,287 | 2,24 |
| x = 0,50 |  | 1,009 | 2,328 | 2,450 | 2,24 |
| x = 0,25 | I4mcm | 0,867 | 2,234 | 2,351 | 2,12 |
| x = 0,0 |  | 0,724 | 2,142 | 2,255 | 2,11 [10] |

It is well known that the PBE method is not accurate enough to calculate the energy of excited electronic states. This fact is reflected by comparing the obtained gaps in Table 6. It can be observed that PBE results are far from the experimental data. On the other hand, the TB-mBJ method provides substantial agreement for gap calculations in a wide variety of systems, such as those studied here. It is observed that, for orthorhombic structures, the gap calculated with the TB-mBJ1 potential is in better agreement with the interpolated experimental values, while those calculated with the TB-mBJ0 potential are closer when the structure is tetragonal. Then, $E_{VB(SCTN)}^{interface} - E_{O2/H2O}^{interface}$ was calculated for each composition using Eq. 6 with both the calculated and the interpolated experimental values for comparison. The results are presented in Table 7.

**Table 7:** Calculated $E_{VB(SCTN)}^{interface} - E_{O2/H2O}^{interface}$ for the CSTN interface in eV, using the gap obtained from interpolated experimental data and calculated with TB-mBJ.

| Space group | Pnma | | | I4mcm | | |
|---|---|---|---|---|---|---|
| Composition | x =1 | x= 0.75 | X= 0.60 | x = 0,50 | x = 0,25 | x = 0 |
| Interpolated | | | | | | |
| $E_{VB(SCTN)}^{interface} - E_{O2/H2O}^{interface}$ | −1,22 | −0,61 | −0,65 | 0,03 | −0,28 | −0,18 |
| Calculated | | | | | | |
| | TB-mBJ1 | | | TB-mBJ0 | | |
| $E_{VB(SCTN)}^{interface} - E_{O2/H2O}^{interface}$ | −1,01 | −0,58 | −0,70 | −0,08 | −0,46 | −0,23 |

Figure 10 illustrates the alignment results obtained using both obtained gaps. In the upper part of the figure (gap obtained by interpolating experimental data), most CSTN compositions result correctly aligned, thus, are suitable for consideration as photoelectrochemical cell materials, except for x = 0.25 in the tetragonal phase, where the CB maximum lies over the O$_2$/H$_2$O level, and possibly x = 1, where the CB minimum is too near the H₂/H₂O redox level. On the other hand, using the TB-mJ0 and TB-mBJ1 method according to the space group of the structures, all compositions meet the necessary alignment condition (Figure 10 lower), unless the composition *x*= 1 that has the same uncertain behaviour. These results correspond with the experimental evidence obtained from Ref [10], providing strong support to the methodology and modelling proposed in this work.

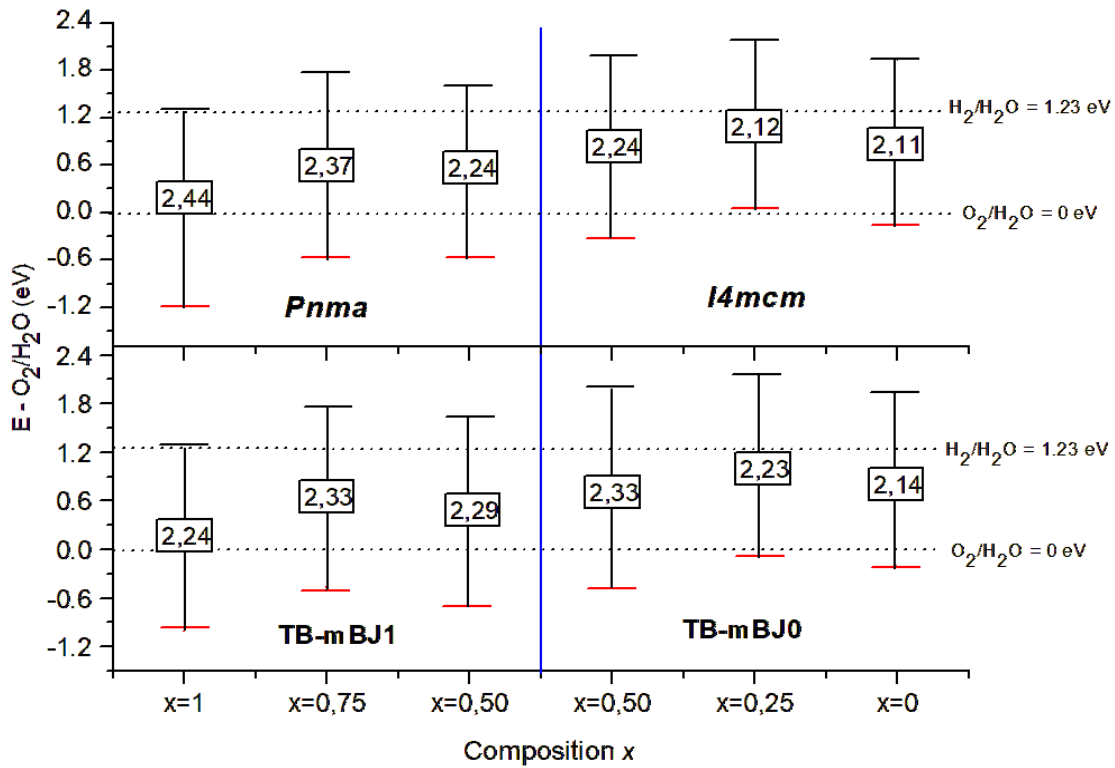

**Figure 10:** Band alignment of the VB and CB (horizontal segments) of CSTN. The horizontal segments in red represent the maximum of the VB and the black ones the minimum of the CB. The dotted lines indicate the redox levels of water H₂O/H₂ and H₂O/O₂. Each vertical segment is labelled inside the rectangle with the gap in eV: upper: alignment obtained with gaps extrapolated from experimental data; lower: alignment obtained with gaps obtained using TB-mBJ1 and TB-mBJ0 method. The blue vertical line separates the orthorhombic and tetragonal compositions.

### 6- Summary and conclusions

Band alignment analysis was carried out on $Sr_{1-x}Ca_xTaO_2N$ solid solutions using first-principles calculations for the compositions *x* = 0, 0.25, 0.5, 0.75, and 1. The non reported structure for *x* = 0.5 near the morphotropic phase transition was energetically analysed, predicting that the *I4mcm* structure is more stable compared to the *Pnma* structure. For the gap determination, the GGA PBE method and two different forms of TB-mBJ were used. It was found that, firstly, the GGA PBE method does not acceptably reproduce the experimental values, as is already known. On the other hand, TB-mBJ results are in good agreement with them. However, TB-mBJ0 reproduces better the gap for the cells with tetragonal symmetry, while TB-mBJ1 does so for those with orthorhombic symmetry.

The band alignment of the solid solutions with the $H_2O$ redox levels was analysed based on the method proposed by Wu et al. [11] considering the band bending effect at the interface. In our approach SC were used, and its minimum height was established in order to avoid possible surface effects. Also, as already detailed in our previous work, the effect of the relaxation of the atoms close to the surface was implemented. Additionally, the height of the water box was designed to avoid periodic boundary conditions effects. The liquid water phase was simulated using both classical and quantum mechanical methods to obtain a realistic representation of the interaction with the semiconductor. For the VB alignment, two approaches were tested. On the one hand, the interpolated gap experimental results were used. On the other, the corresponding TB-mBJ for each composition predicted gaps were employed. The last method resulted in better agreement with the experimental observations. All these factors were crucial to understand the behaviour of the semiconductor surface and its capability to produce water splitting.

The implemented methodology arrives to results that coincide with experimental findings: that most of the studied structures meet the necessary condition required for the material to be suitable for photoelectrochemical applications, except for the structure with $x$ = 1 of pure oxynitride, where the CBM is observed at the edge of the $H_2/H_2O$ redox level. This work shows that isovalent substitution in oxynitrides $ABO_2N$ between Sr and Ca at the A site to form solid solutions significantly contributes to tune and improve band alignment, and providing strong support to the methodology and modelling here proposed to study other oxynitride/$H_2O$ interfaces.

**Acknowledgments.**

This work was partially supported by Consejo Nacional de Investigaciones Científicas y Técnicas (CONICET) under grant PIP11220220100435CO and Facultad de Ciencias Exactas, Universidad Nacional de La Plata under grant X843. For the calculations, we used the computational resources from CCAD – Universidad Nacional de Córdoba (https://ccad.unc.edu.ar/), which are part of SNCAD – MinCyT, República Argentina.